# Stochastic transition model for pedestrian dynamics

Michael Schultz

Department of Air Transport Technology and Logistics,
Faculty of Transport and Traffic Sciences "Friedrich List"
Technische Universität Dresden
schultz@ifl.tu-dresden.de

**Abstract.** The proposed stochastic model for pedestrian dynamics is based on existing approaches using cellular automata, combined with substantial extensions, to compensate the deficiencies resulting of the discrete grid structure. This agent motion model is extended by both a grid-based path planning and mid-range agent interaction component. The stochastic model proves its capabilities for a quantitative reproduction of the characteristic shape of the common fundamental diagram of pedestrian dynamics. Moreover, effects of self-organizing behavior are successfully reproduced. The stochastic cellular automata approach is found to be adequate with respect to uncertainties in human motion patterns, a feature previously held by artificial noise terms alone.

## 1    Introduction

Models for pedestrian dynamics cope with different aspects of behavior of human motion patterns. Generally, such models refer to (one or more of) three levels of motion planning: operational (short range), tactical (medium scale), and strategic behavior (large scale). The basic microscopic models (e.g. social force, cellular automaton, or discrete choice) particularly focus on the operational level of motion planning and execution. Especially, the often favored social force approach (Helbing and Molnar, 1995), which defines attraction and repulsion forces between humans, turns out as a good analogy to reproduce substantial effects of self-organization. In contrast to the social force model (cf. Johansson et al., 2007; Moussaid et al., 2010), the incrementally developed motion model (Schultz et al., 2005, Schultz, 2010) is based on a stochastic cellular automata approach (Burstedde et al., 2001) to handle the unpredictable behavior of individual path deviations. This paper starts out with an overview of the existing model and then focuses on the identified deficiencies in detail to introduce compensation methods that are needed to overcome grid-based influences on the model behavior. Further on, a calibration against the fundamental diagram and a navigational algorithm are presented. The effects of agent-to-agent interactions are finally taken into account by a dynamic motion field.

## 2    Basic Stochastic Model

The developed microscopic model of pedestrian motion is based on a square grid structure (cellular automata) and represents a stochastic approach using a 3x3 transition matrix (Moore neighborhood) for agent movements (Burstedde et al., 2001). In contrast to the common understanding of cellular automata, a paradigm change takes place: instead of changing the cell status depending on the status of the surrounding cells (pull concept), the agent moves over the regular grid and enters cells, which have to be free of other agents or obstacles (push concept). To model the undisturbed agent motion, the velocity vector is divided into independent Cartesian components (longitudinal and lateral), which yields three discrete motion events: forward, center, backward and left, center, right respectively. Each motion component is characterized by a specific motion probability $\mu$ with a corresponding motion variance $\sigma^2$, which result in a generalized transition probability $p$ as defined in eq. (1)-(3).

$$p_{\text{forward or left}} = \tfrac{1}{2} \; \sigma^2 + \mu^2 + \mu \qquad (1)$$

$$p_{\text{center}} = 1 - \sigma^2 + \mu^2 \qquad (2)$$

$$p_{\text{backward or right}} = \tfrac{1}{2} \; \sigma^2 + \mu^2 - \mu \qquad (3)$$

The symmetry of the lateral motion component ($p_{\text{right}} = p_{\text{left}}$) and the assumption that agents do not move backwards ($p_{\text{backward}} = 0$) are simplifying, but reasonable constraints. The motion components are combined to a 3x3 matrix $M_{ij}$ and determine the transition from the center cell to the right adjacent cell (see fig.1 (left), forward motion to $M_{1,0}$).

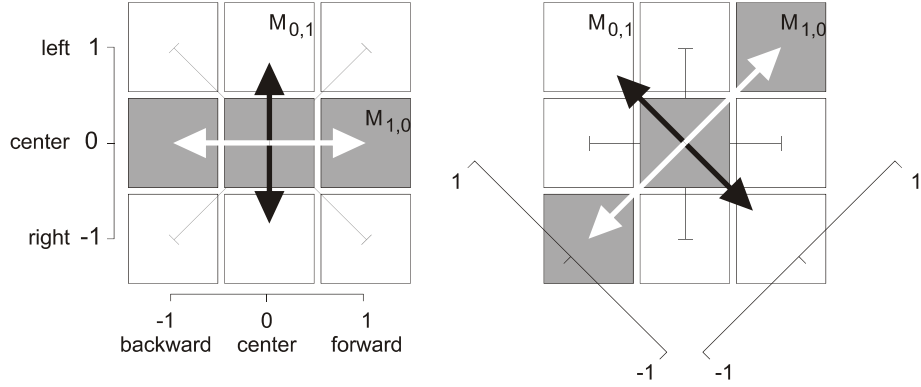

**Fig. 1.** Individual transition to all connected cells with a probability of $M_{ij}$ for the horizontal (left) and diagonal transition (right)

Although the transition matrix apparently possesses a two-dimensional characteristic, only a one-dimensional transition with a lateral deviation (1.5-dimensional) is achieved at this stage. A true two-dimensional transition is ensured by blending two matrices: one matrix represents the horizontal transition $M^{0°}$ (fig. 1, left) and one matrix the diagonal transition $M^{45°}$ (fig. 1, right), which is created by a counterclockwise rotation, whereas the indexes are shifted by one position. In contrast to Burstedde et al. (2001), the specific weights in eq. (4) consider both the motion angle $\alpha$ and the deviating distance of diagonal cells, which additionally results in a normalization of the final transition matrix $M$.

$$M = (1 - \tan \alpha)\ M^{0°} + \tan \alpha\ \sqrt{2}\ M^{45°} \qquad (4)$$

The consideration of the rotational symmetry allows a mapping of all motion directions a $\alpha \in [0, 360°]$ (Burstedde et al., 2001; Schultz, 2010). Three examples of final transition matrices are presented at the following fig. 2, whereas the longitudinal variance ($\sigma_{long.}^2 = \mu_{long} - \mu_{long}^2$) and the lateral transition probability ($\mu_{lateral} = 0$) consequently result from the simplifying constraints. Finally, the parameter set is reduced to $\mu = \mu_{long}$, $\sigma^2 = \sigma_{lateral}^2$, and motion angle $\alpha$.

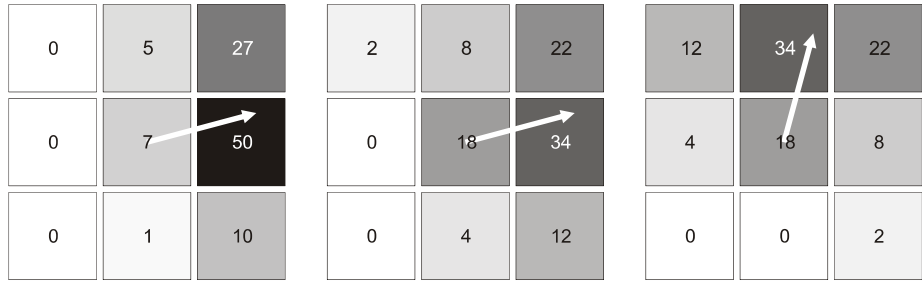

**Fig. 2.** Sample states of transition matrix $M$ (transition probability in %) using the blending function of eq. (4) with specific parameter sets for $\mu$, $\sigma^2$, $\alpha$: {0.9, 0.3, 15°}, {0.7, 0.4, 15°}, and {0.7, 0.4, 75°} (from left to right)

Detailed analysis of this stochastic model points out significant directional dependencies which are caused by the underlying regular grid structure. As an example, entering diagonal cells implies a longer path (similar to a higher speed) in comparison to entering horizontal/vertical located cells. Consequently, the next section focuses on these model deficiencies and introduces reliable compensation strategies.

## 3   Deficiencies and Compensations

Besides the obvious distance discrepancy the blending of the horizontal and diagonal matrices results in additional model deficiencies, which affects the operational motion behavior of the agents. A direct transition to the adjacent cells is only possible with rotation angles in multiples of 45°. All other angles immanently result in a stochastic selection process (fig. 3). This implies both an additional motion variance and a different transition/avoidance behavior. In the case of a horizontal transition (fig. 3, left) an occupied cell (black) leaves no alternatives and will effectively stop the agent while the motion angle of 22.5° yields equal probabilities (using blending function of eq. (4)) and two optional transitions exist (fig. 3, right).

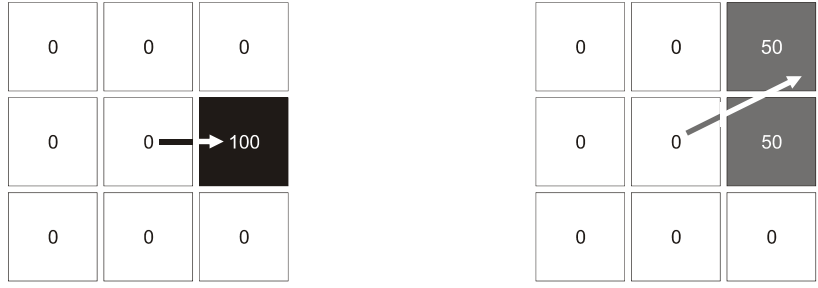

**Fig. 3.** Unique transition to exact one cell with $\mu_{forward} = 1$ and $\sigma_{left/right}^2 = 0$ (left), which is only feasible for transitions at the axes of symmetry, otherwise a cell selection is inevitable (right)

The effect of different distance and variance characteristics is shown in fig. 4; the distance error increases and reaches a maximum at $\alpha = 45°$ and the movement deviation increases and decreases in the range between 0 and 45°. Furthermore, these characteristics are directly coupled and impose a damping effect on each other: increasing lateral variance leads to decreased distance differences and a decreasing probability of forward motion leads to a more adjusted behavior with a smaller variance difference (fig. 4). To harmonize the different transition behavior caused by the motion angle dependencies, specific compensation settings for the transition matrix are numerically calculated. The compensation settings qualitatively points out an inverse characteristics to the identified error, whereas both the cross correlation of $\mu$ and $\sigma^2$ as well as the less smaller influence of additional angle deviations are taken into account (Schultz, 2010).

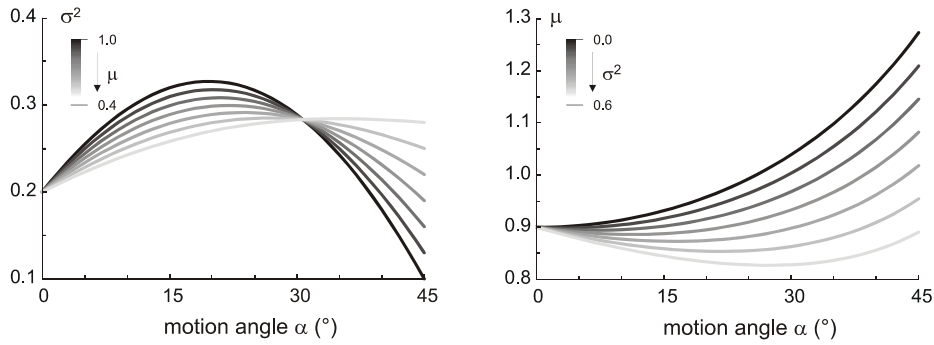

**Fig. 4.** Error characteristics of motion deviation (left) and motion speed (right) depending on the cross correlation regarding to the motion speed ($1 < \mu < 0.4$ at constant $\sigma^2 = 0.2$) and the motion deviation ($0 > \sigma^2 > 0.6$ at constant $\mu = 0.9$) respectively.

## 4 Calibration

To complete the stochastic movement model, additional boundary conditions have to be defined regarding both model and agent behavior:

- the update procedure of simulation environment (agent scheduling),
- the behavior in the case of intersecting movement paths (common collision avoidance strategy),
- the amount of motion per simulation turn (number of steps),
- behavioral implications of blocked cells (obstacles, other agents), and
- temporarily blocking occupied cells during the movement phase (motion trace).

The sequential update procedure is used in the model, since its points out no significant drawbacks, if the sequence is shuffled at each simulation step (Klüpfel, 2003; Wölki, 2006; Schultz, 2010). Thus, the constellation of parallel access to one free cell will be avoided (fig. 5, left). The influences of crossing paths (fig. 5, center and right) are directly related to the last condition of considering traces and since this trace condition possesses a significant influence to the model, the crossings are not permitted. To evaluate the relevance of the amount of motion per simulation turn, the fundamental diagram is used as a commonly excepted reference for the speed-density relation of pedestrian dynamics (Weidmann, 1992). To allow a reliable scenario evaluation an infinite environment was set up with:

- varying amount of motion per simulation turn (from one to five motion steps),
- always-move vs. waiting procedure at blocked cells, and finally
- the ability of agents to leave a trace of blocked cells during a simulation turn.

The always-move procedure is implemented by the condition that each blocked cell relatively increases the transition probability of the free cells in equal measure to their specific transition probability (only free cells could be selected). The waiting procedure implies that agents immediately stop their motion at the current simulation turn if a blocked cell is stochastically selected for the transition.

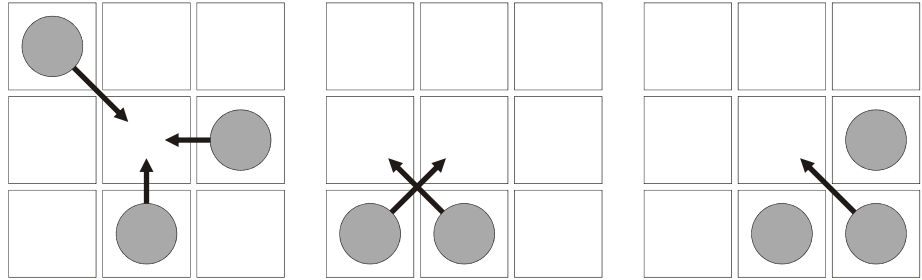

**Fig. 5.** Demand for determination of movement rules to cope with parallel transition to one free cell (left) and crossing paths (center and right)

The determined movement rules in connection with the ability of motion traces (temporally blocked cells) results in the associated speed characteristics as shown in fig. 6. The left diagram of fig. 6 (waiting rule) points out no clear tendency to consider a specific amount of steps per simulation turn. But the always-move rule successfully matches the speed-density relation of pedestrian dynamics assuming 3-4 steps at one simulation turn. (fig. 6, right).

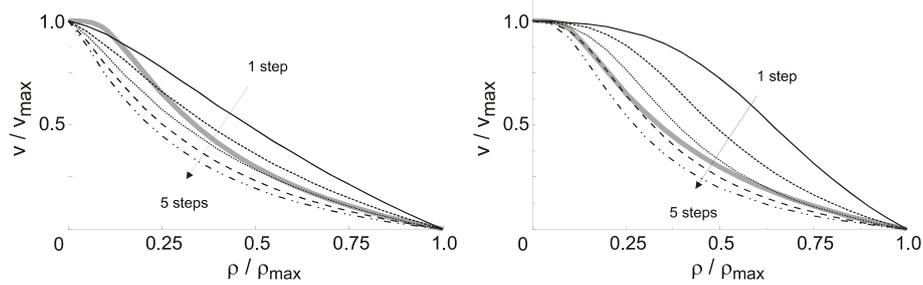

**Fig. 6.** Normalized speed-density characteristics considering one to five motion steps per simulation turn as well as motion trace for both waiting rule (left) and always-move rule (right) against the common characteristic of pedestrian dynamics (bold gray line, according to Weidmann (1992))

Up to this point, the stochastic motion approach only considers the surrounding neighbors, which are directly connected to the local cell position. In comparison to the fundamental diagram of pedestrian dynamics (correlation of individual motion speed and agent density), the developed model reproduces the expected shape if the agent:

- will not wait if other agents block his way (occupied cells increase the transition probability of surrounding cells),
- moves three/four steps at one simulation turn (depending on agent density), and
- leaves a motion trace (temporally block all cells entered at the current simulation turn).

With increasing agent density ($\rho/\rho_{max} > 0.6$) the number of steps per simulation turn should be reduced from four to three. At fig. 7 the adequate characteristics of the stochastic movement model is represented in detail. It needs to be emphasized, that low density scenario ($\rho/\rho_{max} < 0.3$) significantly benefits from a four step implementation whereas the three step implementation leads to slight overestimations compared to the common speed-density relation of the pedestrian dynamics. A normalized notation of this relation is given by eq. (5), with $v_{max} = 1.34$ m/s, $\rho_{max} = 6.25$ pedestrian/m$^2$, and $\rho \in (0, \rho_{max}]$.

$$\frac{v}{v_{max}} = 1 - e^{0.3061\left(1 - \frac{\varrho_{max}}{\varrho}\right)} \tag{5}$$

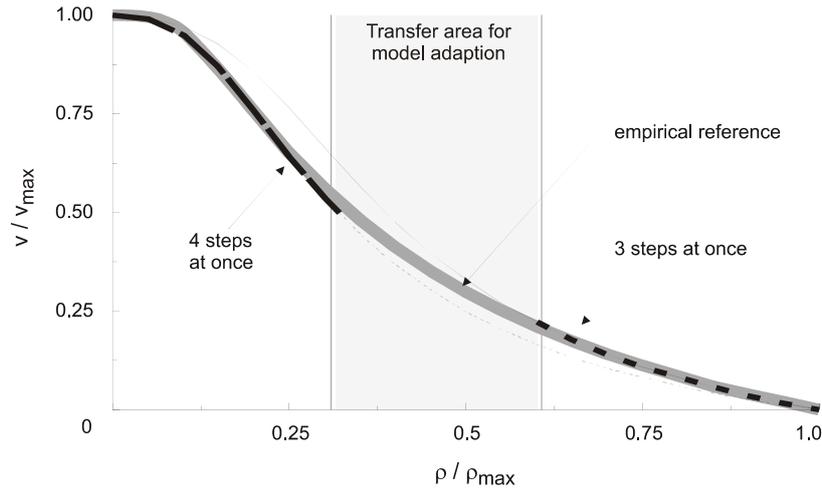

**Fig. 7.** Detailed view to the characteristic of the developed modal against the fundamental diagram of pedestrian dynamics (Weidmann, 1992)

## 5   Navigation

For modeling the motion of pedestrians, two main approaches have been evaluated during the least three decades: one uses a set of navigation points to navigate agents around obstacles (de Berg, 1997), the other one relies on a discretized grid structure which holds guidance information (walking distance to destination, the direction of shortest/ quickest path) and efficiently takes obstacles into consideration as well. The most prominent and most widely used method that calculates Euclidean distances is a numerical solution of the Eikonal equation (Bruns, 1895; Frank, 1927). The Fast Marching Method (FMM) is a comparatively fast method and possesses a comparatively small deviation regarding to the Euclidean distance (Osher, 1988; Sethian 1996, 1999; Kimmel 1998; Hartmann, 2010). The FMM shows an optimal worst-case behavior (concerning the computation time) and the relative distance error in general decreases with the increasing distance from the destination, implying that a finer grid reduces the distance error. Apart from optimizing the numerical Eikonal equation solver algorithmically, simplified methods could trade calculation exactness for computation speed. The most noticeable example for this is a simple flood fill algorithm over common edges/corners, resulting in p-norm metrics for the vector x (see eq. (6)), with $p = \{1, 2, \infty\}$ for the Manhattan, Euclidean, and Chebychev metric respectively.

$$\|x\|_p := \left( \sum |x_i|^p \right)^{1/p} \qquad (6)$$

The flood fill algorithm (breadth-first search) for the creation of the potential field was introduced by Lee (1961) for the von Neumann neighborhood and is consequently applied for the Moore neighborhood (Schultz, 2010; Schultz et al., 2010).

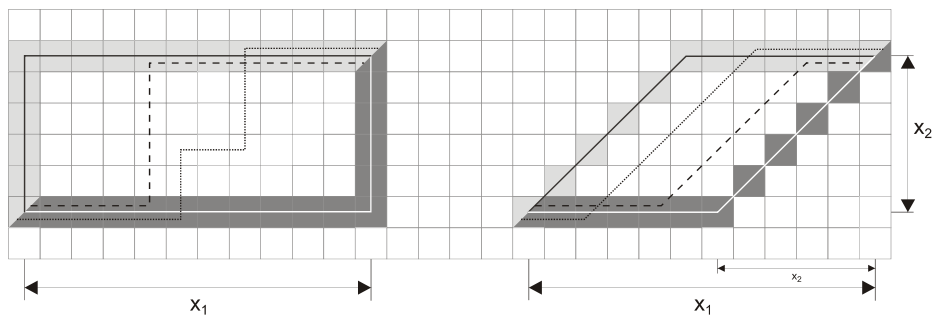

**Fig. 8.** Manhattan metric (von Neumann neighborhood, left) and distance metric grid considering a square grid and a connection over common corners (Moore neighborhood, right); dotted lines represent equivalent motion paths (same length) and the colored cells are the boundaries of these valid paths

With the Euclidean metric as correct solution, these methods lead to relative errors which remain constant over distance or even increase, which means that a finer grid size does not necessarily improve the precision arbitrarily (fig. 9).

However, it is possible to reduce the error by making some slight modifications upon the simple flood fill methods (Kretz, 2010). In comparison to the Manhattan (quadrant I, see fig. 9), Euclidean (quadrant II) and Chebychev (quadrant III) metrics, the used flood fill approach for the Moore neighborhood possess a distance metric (quadrant IV) as determined in eq. (7).

$$x := \Delta x + \overline{2} \min x_i \tag{7}$$

Although the flood fill algorithm obviously features a closer approximation of the Euclidean distance it tends to overestimate the distances.

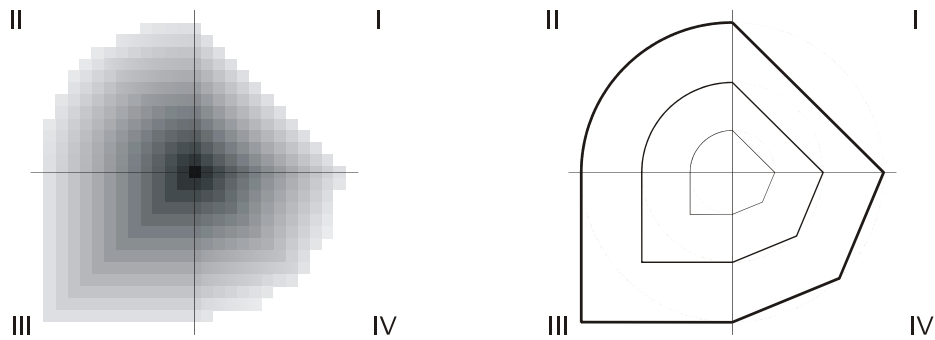

**Fig. 9.** Comparison of Manhattan (I), Euclidean (II), Chebychev (III) metrics against the derived Moore metric using a color-coded distance field (left) and equidistance lines (right)

Regardless of the non-conforming approximation of distance, the Moore metric possesses a significant characteristic which may be used to determine the correct walking direction with respect to the shortest way (Schultz, 2010; Schultz et al., 2010). This method does neither rely on a computation time expensive sorting of distances of currently *active* cells as the FMM does, nor does it even has the need of the calculation of square roots or similar rather computation-time expensive functions.

As an elementary example an obstacle configuration is set up to calculate the direction from each position of the environment to the highlighted circle (fig. 10, left). The calculated distance is outlined as a color-coded distance image (black: distance = 0, white: distance $\to \infty$; fig. 10, right). A closer view on this image highlights the exposed characteristics of the axis of symmetry (cf. fig. 11). If agents would use this distance field according the rule to enter only adjacent cells with the smallest distance value, finally their movements will end up at these axes which results in unnatural crowed areas (jam from nowhere).

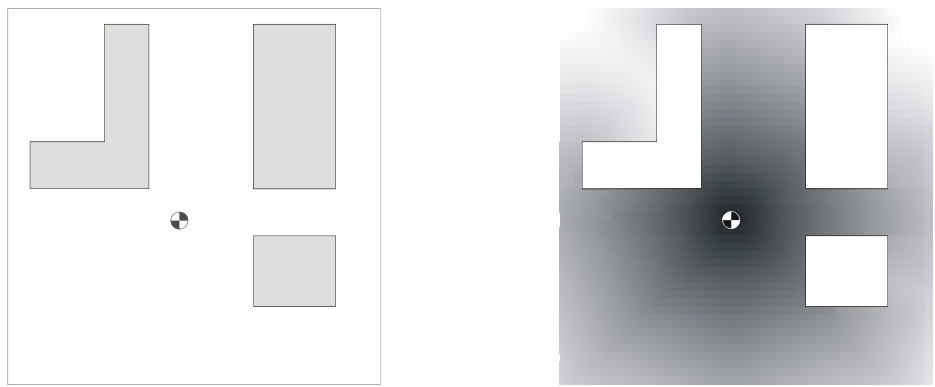

**Fig. 10.** Sample environment with centered destination (circle) and obstacles (rectangles) (left) and the calculated, color-coded distance field using the Moore metric (right)

In contrast to common smoothed gradient procedures to derive the local motion vectors, the developed algorithm creates the motion vector field already during the flood fill process by storing the direction of the downstream position (originating cell position). The flood fill algorithm starts at the target cell and stores the distance at each surrounding cell, if the already stored distance value is greater than the current distance value. If all adjacent cells filled with the distance information, these cells are used as the next source of distance information. So each cell possesses an originating cell, which was the base for the own distance information. The location of this downstream cell depends on the processing sequence of the algorithm. The idea of distance sorted cells to ensure short computation times by preventing multiple calculations of cell was tested without any indication of potential improvements. But a detailed verification of both clockwise and counter-clockwise handling of the cell calculation points out significant directional patterns. Due to the fact that each cell has two possible downstream cells, one diagonal cell and one horizontal/vertical cell (neglecting obstacles), the handling sequence of the flood fill algorithm determines which cell is selected first. As a result, the counter-clockwise cell handling leads to an opposite selection as the clockwise cell handling sequence (fig. 11). Each cell handling sequence promotes one specific pattern depending on the direction angle to the destination. Fig. 11 clearly shows the both directional patterns (cf. lower left quadrant with opposed upper and lower triangles) and a counter-/clockwise behavior.

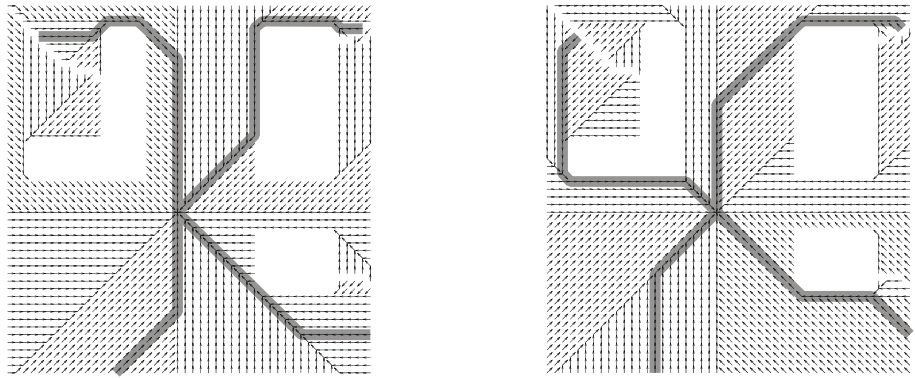

**Fig. 11.** Contrary heading vector fields based on a distance field calculated clockwise (left) and counter-clockwise (right)

The combination of both cell handling procedures leads to a valid solution of the movement vector field once they are combined to one aggregated field. Fig. 12 shows the aggregation process using four generic starting positions at the outer corners of the quadrants. The common rhombus shape of each single distance segment is emphasized (cf. fig.8, right). Addressing the geometric definition of the rhombus, two vector components are used (derived from diagonal and horizontal/vertical connections). The complete directional grid is shown in fig. 12 (right) and indicates no directional artifacts and the declared paths are fully equivalent to the Euclidean distance metric.

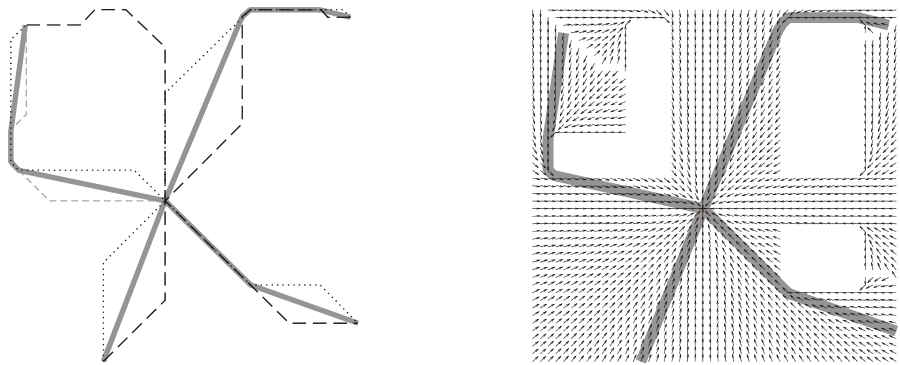

**Fig. 12.** Combination of clock- and counterclockwise vector fields to a valid directional grid equivalent to the Euclidean distance metric (shortest distance)

## 6    Interactions

The calculation of directional fields for motion destinations is an essential part of each pedestrian motion model and have to consider several environmental conditions. The proven quality of the algorithm underlines the relevance of the achieved results to the navigational tasks of the agents, even considering static obstacles (Schultz, 2010; Schultz et al., 2010). In contrast to the static obstacles, dynamic objects are not taken into account on the path planning process (tactical movement level). Although the introduced stochastic motion model considers individual traces during the movement updates, these effects are only confined to a small area (three/four steps, as defined in section 4). Against the background of game design, the interactions of agents are an essential strategic behavioral component and basically considering spheres of influence (Zobrist, 1969) which is one basic technique (influence maps) for tactical movement decisions of agents (Pottinger, 2000; Sweetser, 2004; Tozour, 2001, 2004; Woodcock, 2002). A commonly accepted approach to include tactical movement decisions in cellular automata models for pedestrian dynamics is the calculation of a dynamic floor field (Burstedde, 2001; Schadschneider, 2002).

Using the idea of the repulsive impact of the walking path of other agents to the individual movement decision a dynamic motion field will be introduced. This grid combines both future position based on the current motion direction and past positions (walking paths). In contrast to the potential field, the cells store movement vectors, where the different characteristic of possible future and past position are considered. Each entered cell stores the local motion direction of the corresponding agent (fig. 13, left).

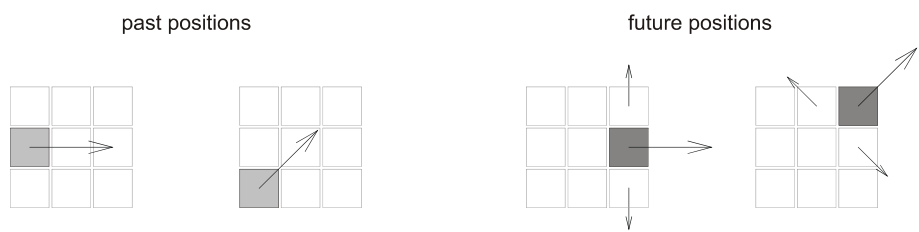

**Fig. 13.** Past positions and future positions are stored in a dynamic motion field, whereas future positions (path extrapolation) features both an additional lateral component and a slight different characteristics (cell position) for diagonal transitions

Further on, the current local motion direction is used to mark potential future cells in advance (cells are derived from the extrapolation of the prior simulation turn), whereas the adjacent lateral cells store the orthogonal motion vector (fig. 13, right). This procedure reflects the observed motion behavior of pedestrians to avoid crossed movement paths and to follow preceding pedestrians. Finally, one simulation turn consists of four sub-steps:

1. Randomly sequenced agents update their position using the direction to their individual destinations superposed by the local vector of the dynamic motion field.
2. The local motion direction is added to the dynamic motion field for each entered cell.
3. Considering the current position, the motion steps during the simulation turn are used to extrapolate the motion path.
4. Finally, the entire dynamic motion field is smoothed by a convolution filter using the one-dimensional kernel ($dc/3$, $dc/3$, $dc/3$) with $dc$ [0, 1] as the decay coefficient.

The two step application of the kernel in horizontal and vertical grid direction leads to an isotropic Gaussian filter behavior (see fig. 14 from left to right).

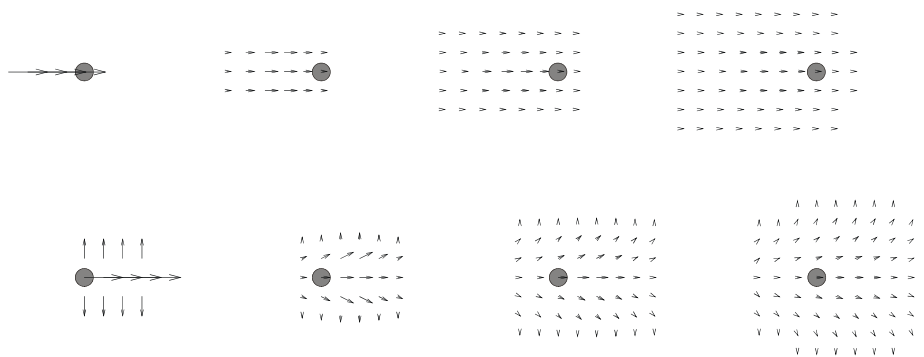

**Fig. 14.** Dynamic motion field to mark past positions (top) and future positions (bottom), starting with the initial field on the right considering an isotropic expansion associated with a temporally decay caused by a Gaussian filtering process

In fig. 15 the result of a complete walking path is shown, considering past/future positions, isotropic spatial diffusion, and temporally decay behavior.

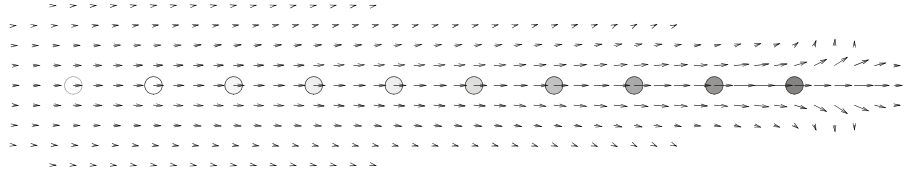

**Fig. 15.** Combined dynamic field of agent motion to cover individual movements of a certain period of time

The proposed method of a dynamic motion field can also be applied for semi-static obstacles, such as waiting agents, small objects (e.g. baggage, signs, park bench), or temporally blockings of the infrastructure. In contrast to moving agents, the initial vector field of the semi-static obstacles possesses a symmetric characteristic (fig. 16, left). The repulsive effect of non-moving objects increases if these objects are closely located to each other. In fig. 16 (right) the superposition of the destination-directed motion field (one-sided walkway from left to right) with the dynamic motion field of the semi-static obstacles evidently points out a collision avoiding behavior. It may be noticed, that the presented example is only a proof of concept and a detailed parameterization has to be done at the field.

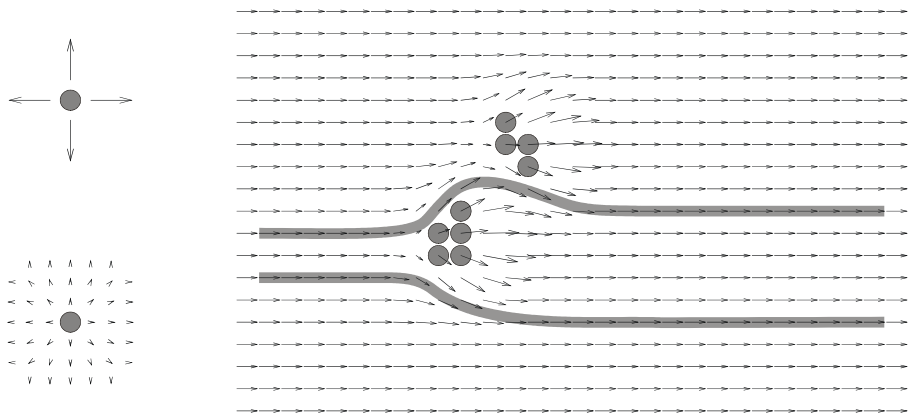

**Fig. 16.** Dynamic motion field for static objects (upper left), isotropic expanded with a temporally decay (lower left), and a convincing example of the combination of movement direction field and individual vector fields (right)

The simulation result of a counter-flow scenario is shown in fig. 17, whereas the separation of agents regarding their movement direction is clearly indicated. Agents with opposite headings avoid collisions considering the dynamic motion field and downstream agents use the trace information to follow agents with the similar heading.

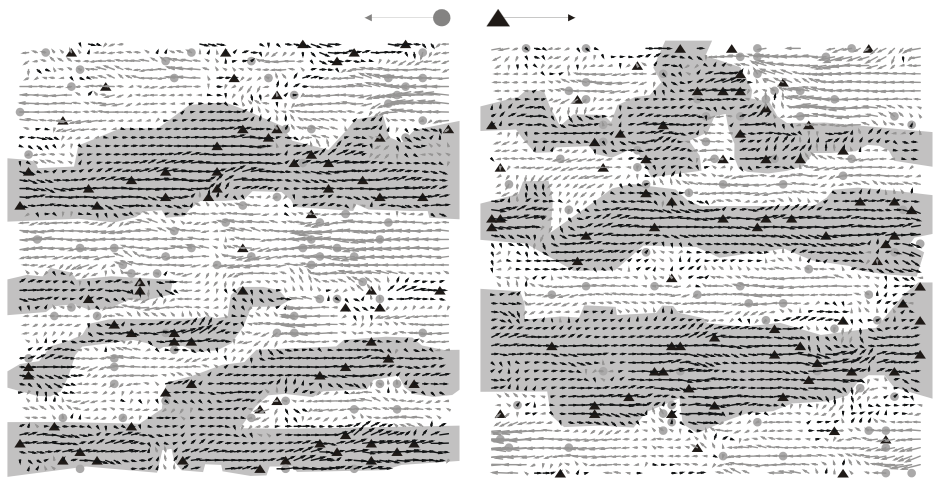

**Fig. 17.** Result of the complete stochastic model for pedestrian dynamics using a counter flow scenario: sample lane formations with locally aligned agents

Over the course of time lane formations occur, characterized by areas with agents possessing a comparable motion direction. Due to the fact, that the proposed motion model is based on an immanent stochastic approach, these lanes are not stable in time and space.

# 7    Summary

The developed model is a substantial extension of a stochastic cellular automata approach. The model development is completed by adding agent-oriented environment analysis, route planning, and mid-range agent interaction. The stochastic motion model proves its capabilities for a quantitative reproduction of the characteristic shape of the common fundamental diagram of pedestrian dynamics. The proposed model possesses a $v_{max}$ = 4 (3) speed characteristic, a simple exclusion statistic and a shuffled sequential update procedure. The underlying regular grid structure of the cellular automata results in a direction-dependent behavior of motion speed and variance. The identified model deficiencies are compensated on a fundamental level to ensure valid local transition matrices for the stochastic motion behavior.

The motion direction is derived from a destination-oriented motion field. This field is calculated by a flood fill algorithm using the underlying regular grid considering a Moore neighborhood, where obstacles are represented by occupied cells. The applied algorithm exactly reproduces the shortest path equivalent to the Euclidean distance. The developed algorithm efficiently prevents distance errors and directional artifacts by combining two complementary directional fields, which are based on a Moore distance metric calculation. Due to the fact, that flood fill algorithms are often used in agent-based simulations to create distance potentials for navigational purposes, this enhanced algorithm provides a fundamental contribution to existing and future simulation environments.